\documentclass[12pt,preprint]{aastex}

\slugcomment{Submitted to AJ, 2006 October 09}

\shorttitle{GRB 050813}

\shortauthors{Ferrero et al.}

\received{}

\begin{document}

\title{Constraints on an optical afterglow and on supernova light \\ 
following the short burst 
GRB 050813\thanks{Based on observations collected at the
European Southern Observatory, La Silla and Paranal, Chile  (ESO
Programme 075.D-0415) and on observations taken at the German-Spanish Calar
Alto Observatory and at IAA's Observatorio de Sierra Nevada in Spain.}}

\author{
P.~Ferrero\altaffilmark{1},
S.~F.~Sanchez\altaffilmark{2},
D.~A.~Kann\altaffilmark{1},
S.~Klose\altaffilmark{1},
J.~Greiner\altaffilmark{3},
J.~Gorosabel\altaffilmark{4},
D.~H.~Hartmann\altaffilmark{5},
A.~A.~Henden\altaffilmark{6},
P.~M\o ller\altaffilmark{7},
E.~Palazzi\altaffilmark{8},
A.~Rau\altaffilmark{9},
B.~Stecklum\altaffilmark{1},
A.~J.~Castro-Tirado\altaffilmark{4},
J.~P.~U.~Fynbo\altaffilmark{10},
J.~Hjorth\altaffilmark{10},
P.~Jakobsson\altaffilmark{11},
C.~Kouveliotou\altaffilmark{12},
N.~Masetti\altaffilmark{8},
E.~Pian\altaffilmark{13},
N.~R.~Tanvir\altaffilmark{14},
R.~A.~M.~J.~Wijers\altaffilmark{15}
}


\altaffiltext{1}{Th\"uringer Landessternwarte Tautenburg, Sternwarte 5,
D--07778 Tautenburg, Germany}

\altaffiltext{2}{Centro Astron\'{o}mico Hispano Alem\'{a}n de Calar Alto, 
Calle Jesus Durban Remon 2-2, E--04004 Almer\'{\i}a, Spain}

\altaffiltext{3}{Max-Planck-Institut f\"ur Extraterrestische Physik, 
Giessenbachstra\ss{}e, D--85741 Garching, Germany}

\altaffiltext{4}{Instituto de Astrof\'{\i}sica de Andaluc\'{\i}a (IAA-CSIC), 
Apartado de Correos, 3.004, E--18.080 Granada, Spain}

\altaffiltext{5}{Clemson University, Department of Physics and Astronomy, 
Clemson, SC 29634-0978, USA}

\altaffiltext{6}{U. S. Naval Observatory/Universities Space Research Association, 
Flagstaff Station, Flagstaff, AZ 86001, USA}

\altaffiltext{7}{European Southern Observatory, Karl Schwarzschild-Strasse 2, 
D--85748 Garching bei M\"{u}nchen, Germany}

\altaffiltext{8}{INAF -- Istituto di Astrofisica Spaziale e Fisica Cosmica di Bologna,
via Gobetti 101, I-40129 Bologna, Italy}

\altaffiltext{9}{Division of Physics, Mathematics and Astronomy, 105-24, 
California Institute of Technology, Pasadena, CA 91125}

\altaffiltext{10}{Dark Cosmology Centre, Niels Bohr Insitute, University of 
Copenhagen, Juliane Maries Vej 30, 2100 Copenhagen, Denmark}

\altaffiltext{11}{Centre for Astrophysics Research, University of Hertfordshire, College Lane,
Hatfield, Herts, AL10 9AB, UK}

\altaffiltext{12}{NSSTC, SD-50, 320 Sparkman Drive, Huntsville, AL 35805, USA}

\altaffiltext{13}{INAF -- Osservatorio Astronomico di Trieste, Via Tiepolo 11, 
I-34143 Trieste, Italy}

\altaffiltext{14}{Department of Physical and Astronomy, University of Leicester,
Leicester, LE1 7RH, UK}

\altaffiltext{15}{University of Amsterdam, Kruislaan 403, 1098 SJ Amsterdam, The Netherlands}

\begin{abstract}

We report early follow-up observations of the error box of the short burst
050813 using the telescopes at Calar Alto and at Observatorio Sierra Nevada
(OSN), followed by deep VLT/FORS2 $I$-band observations obtained under very
good seeing conditions 5.7 and 11.7 days after the event. Neither a fading
afterglow, nor a rising SN component was found, so the potential GRB
host galaxy has not been identified based on a comparison of the two VLT images
taken at different epoches. We discuss if any of the galaxies 
present in the original 10 arcsec XRT
error circle could be the host. In any case, the optical afterglow of GRB
050813 was of very low luminosity. We conclude that all these
properties are consistent with the binary compact merger hypothesis for the
progenitor of GRB 050813.

\end{abstract}

\keywords{Gamma rays: bursts: individual: GRB 050813 --- Supernovae: general}

\section{Introduction}

\subsection{Short Bursts}

Much progress is currently being made toward understanding the nature of  the
progenitors responsible for the class of short-duration, hard gamma-ray bursts
\citep[][ see also Appendix B]{Kouveliotou1993}. While the physical link
between long-duration, soft gamma-ray bursts and the core collapse of massive
stars \citep[e.g.,][]{Paczynski1998} has been conclusively confirmed by the
spectroscopic detection of supernova (SN) light following some bursts \citep[
for a review]{Stanek2003,Hjorth2003,Pian2006,Woosley2006}, the nature of the
sources responsible for short bursts remains to be revealed in full. Although
there is a developing consensus in the community that at least some short
bursts are due to merging compact stellar objects
\citep[cf.][]{FWH1999,Aloy2005,Rosswog2005,Oechslin2006,Faber2006}, an
unambiguous observational verification of this model is not an easy task and
has not yet  been accomplished. Furthermore, the origin of a certain fraction
of short bursts as giant flares of magnetars in nearby galaxies seems to be
possible as well \citep[cf.][]{Tanvir2005}. Indeed, the short-hard burst
051103 detected by the Interplanetary Network \citep{Golenetskii2005} might be
the first well-localized member of this class \citep{Frederiks2006, Ofek2006}.

Within the context of the merger model, the stellar populations underlying
short bursts could be associated either with an old stellar population or even
with a young one \citep{Belczynski2006}. Short bursts might therefore occur in
quiescent ellipticals or star-forming galaxies.  Indeed, the first short burst
well-localized by \emph{Swift}, GRB 050509B \citep{Gehrels2005}, was
associated with a giant elliptical galaxy located in a cluster of galaxies at
$z=0.225$ \citep{Bloom2006a, Pedersen2005}, while the \emph{HETE-2} short
burst GRB 050709 \citep{Hjorth2005b} occurred in an isolated, star-forming
dwarf galaxy. Shortly thereafter GRB 050724 was found in association with a
lone early-type galaxy
\citep{Bloom2005,Prochaska2005,Berger2005a,Gorosabel2006}.  Assuming as a
working definition that a short burst should have $T_{90}<2$ s, then since GRB
050813 six further short bursts have been accurately localized by
\emph{HETE-2} or \emph{Swift} via their X-ray afterglows by the end of
September 2006 \citep[see also table 8 in][]{Donaghy2006}. Among them GRB 051210
\citep{LaParola2006}, GRB 060502B \citep{Bloom2006b} and GRB 060801
\citep{Racusin2006} had only X-ray afterglows, while GRB 051221A
\citep{Soderberg2006},  GRB 060121 \citep{Malesani2006,Levan2006,dUP2006} and
GRB 060313 \citep[ Hjorth et al. 2007, in preparation]{Roming2006} have
detected optical afterglows as well. A broad range of morphological types of
host galaxies was derived for this set. For example, \citet{Bloom2006b}
postulated an  association between GRB 060502B and a bright elliptical galaxy
at a large offset at $z=0.287$, while GRB 051221A is associated with an
isolated star-forming dwarf galaxy \citep{Soderberg2006}, and the host of GRB
060121 might be a dusty edge-on irregular or spiral galaxy
\citep{Levan2006}. This ``mixed-bag'' of host types is consistent with the
idea that merging compact binaries will sample all types of galaxies, even
those in which star formation turned off a long time ago.  The short burst GRB
050813 belongs to the small set of short bursts for which up to date it has
not been possible to define  precisely the host galaxy.

\subsection{GRB 050813}

According to its observed duration ($T_{90}$, see below), GRB 050813 can be
associated with the class of short bursts with very high (99.9\%) probability
\citep{Donaghy2006}.  In addition, its measured spectral lag is consistent
with zero, another important property  of short bursts \citep{NB2006,
Donaghy2006}.  Furthermore, the small original \emph{Swift} XRT error circle
encompasses parts of an anonymous cluster of galaxies with ellipticals inside
and close to the error circle
\citep{Gladders2005,Gorosabel2005,Prochaska2006}. Taken together,  these
observations suggest that  GRB 050813 should be considered as a typical short
burst.

GRB 050813 was detected by the \emph{Swift} satellite on 2005 August 13,
6:45:09.76 UT \citep{Retter2005}. Its duration in the 15-350 keV band was
$0.6\pm0.1$ seconds \citep{Sato2005}, making it after GRB 050509B and 050724
the third short burst that \emph{Swift} localized quickly and precisely.  It
is reminiscent of GRB 050509B, which had a very faint X-ray afterglow
\citep{Gehrels2005}.  Ground analysis of the X-ray data revealed a faint,
uncatalogued source at coordinates RA, DEC (J2000) = 16$^{\rm h}$ 07$^{\rm m}$
57$\fs$0, +11$^{\circ}$ 14$'$ 52$''$ with an uncertainty of 10 arcsec radius
\citep{Morris2005}. This position was later refined by \cite{Moretti2006} to
RA, DEC (J2000) = 16$^{\rm h}$ 07$^{\rm m}$ 57$\fs$07, +11$^{\circ}$ 14$'$
54$\farcs$2 with an uncertainty of 6.5 arcsec radius; an even smaller error
region was reported by \cite{Prochaska2006}. No optical or near-infrared
afterglow candidate was found. \cite{Li2005} reported an unfiltered upper
limit of magnitude 18.6 at 49.2 seconds after the burst. UVOT observations
started 102 seconds after the trigger and a 3-sigma upper limit of $V=19.1$
was derived from a  188 seconds exposure \citep{Blustin2005}. 
\cite{Sharapov2005} found a limiting  $I$-band magnitude
of $\sim$21 at 10.52 hours after the burst, while \citet{Bikmaev2005} reported
an $R$-band upper limit of $\sim$23 at 12.75 hours after the event.

Spectroscopy of galaxies close to and inside the XRT error circle revealed a
mean redshift of $z=0.72$  \citep{Berger2005b,Foley2005,Prochaska2006},
indicating the possibility that this may also be the redshift of the GRB. This
was later refuted by \cite{Berger2006}, who argued that the host is a 
background galaxy at a (photometric) redshift of about 1.8, possibly
related to a background cluster of galaxies. This would make GRB 050813 the
second most distant \citep[after GRB 060121,][]{dUP2006,Levan2006} 
short burst for which a redshift could be estimated.

Here we report on a  deep follow-up observing campaign of GRB 050813 with
telescopes  at Paranal, Chile, as well as at Calar Alto and at the
Observatorio Sierra  Nevada (OSN), Spain. The constraints we can set on any SN
component following  this burst as well as the faintness of its optical
afterglow  match well into what is known so far about the properties of short
bursts.   Throughout this paper we adopt a world model with  $H_0=71$ km
s$^{-1}$ Mpc$^{-1}$, $\Omega_{\rm M}=0.27$, $\Omega_\Lambda=0.73$
\citep{Spergel2003}, which for  $z$=0.72 yields a distance modulus of 43.22
mag. The luminosity distance is $1.36\,\times\,10^{28}$ cm and  1 arcsec
corresponds to 7.23 kpc. If $z$=1.8, the corresponding numbers are 45.7 mag,
$4.26\,\times\,10^{28}$ cm, and 8.55 kpc.

\section{Observations and data reduction}

A first imaging of the GRB error box was performed with the 1.5-m OSN
telescope at Observatorio Sierra Nevada and the Calar Alto 2.2-m telescope
equipped with CAFOS starting already 0.5 days after the burst
\citep{Gorosabel2005}.  Unfortunately, these observations resulted only in
upper limits for the magnitude of any optical transient (Table~\ref{log}).  In
order to set constraints on a rising SN component, we  have then carried out
deep follow-up observations using VLT/FORS2 in  standard resolution (SR)
imaging mode with a scale of 0.25 arcsec per pixel (field of view
6$\farcm$8 $\times$ 6$\farcm$8). Observations were performed in the
Bessel $I$ band in order to minimize the potential influence of host
extinction on the discovery of a fading (afterglow) or a rising (supernova)
source. A first run was performed on August 19.061 to 19.088 UT, 5.8 days
after the burst. Ten frames were obtained, 200 seconds exposure time
each. Seeing conditions were very good, $\sim$ 0.5 arcsec. A second run using
the same instrumental setup was performed on August 24.990 to 25.017 UT, 11.7
days after the burst. Atmospheric seeing conditions were even better than
during the first observing run, approaching 
0.35 arcsec. Both nights were photometric.

The FORS2 images were bias-subtracted and flat-fielded with standard reduction
procedures provided within IRAF.\footnote{http://iraf.noao.edu} Frames
obtained on the same night and in the same band were summed together in order
to increase the signal-to-noise ratio. Photometry was performed with standard
Point Spread Function (PSF) fitting using the DAOPHOT II image data analysis
package "PSF-fitting\footnote{The PSF-fitting photometry is accomplished by
modeling a two-dimensional Gaussian profile with two free parameters (the half
width at half maxima along $x$ and $y$ coordinates of each frame) on at least
five unsaturated bright stars in each image.}  algorithm" \citep{Stetson1987}
within the MIDAS
platform.\footnote{http://www.eso.org/projects/esomidas} In addition, we
performed aperture photometry using the IRAF Aperture Photometry Package
Apphot.

Additional spectroscopic observations covering the entire original  $r$=10
arcsec XRT error circle \citep{Morris2005} were performed with the Integral
Field Unit VIMOS/IFU at the ESO-VLT starting 20 hours after the burst.
Unfortunately, these observations could not be implemented  into this study
due to technical problems with the data.

Figure~\ref{field} shows the \emph{Swift} XRT 90\% containment radius reported
by \cite{Morris2005} (large circle), the refined error circle by
\cite{Moretti2006} (small circle) and, as a small ellipse, the re-analyzed
X-ray error box (68$\%$ containment radius) given by  \cite{Prochaska2006}. In
the original $r$=10 arcsec XRT error circle we identify 11 sources, designated
by the letters C, D, E, F and the numbers from 1 to 7. Note that B = X, C = B,
4 = B* and E = C in the nomenclature of \cite{Prochaska2006}. The X-ray error
box published by  \cite{Prochaska2006} contains only two sources,  of which
\#6 is the one identified by \cite{Berger2006} as the possible host galaxy
possibly related to a cluster  of galaxies\footnote{E. Berger, talk  given at
``Swift and GRBs: Unveiling the Relativistic Universe'', San Servolo, Venice
(Italy), 2006 June 5-9} at $z$=1.8.   Nothing can be said at this stage about
the redshift of source \#7, however. Here, we assume that it is  a member of
the cluster of galaxies at $z$=0.72 
\citep{Berger2005b,Foley2005,Prochaska2006}.

\section{Results}

Our two FORS2 observing runs were arranged such that they would allow us to
search for a fading (afterglow) as well as for a rising (supernova) component
following GRB 050813, supposing $z$=0.72. Initially we searched for a
transient isolated point source in the  original 10 arcsec XRT error circle,
but we did not find one. The fact that the sources \#2, \#5 and \#6
(Fig.~\ref{field}; Table~\ref{tab2}) are not detected in the combined image of
the first VLT/FORS2 observing run might be due to the presence of the Moon,
causing an enhanced sky background level. During the second FORS2 run the sky
background was much lower and the seeing even better than during the first
observing run. We conclude that any well-isolated afterglow or supernova in
this field was fainter than the magnitude limits at the time of the two FORS2
observing runs, $I$=25.1 and 25.5, respectively.

\subsection{Search for a fading afterglow component}

Based on our deep FORS2 observing runs,  we searched for a potential fading
afterglow superimposed on the brightest extended sources (galaxies) in the
field (Table~\ref{tab1}). No evidence for variability due to an underlying
transient source was found. \cite{Prochaska2006}  identified object C and E as
elliptical galaxies (Fig.~\ref{field}), with C being the most likely host
candidate based on its location relative to their revised elliptical error
circle. In our images source  E appears to have an irregular halo which does
not support its classification as an elliptical. Image subtraction did not
reveal any transient source superimposed on this galaxy. 

In order to obtain an upper limit on a possible detection of an afterglow 
(or a SN) in the first (second) epoch FORS2 image superimposed source E, we 
artificially added point sources of different magnitudes to E and then
performed an aperture photometry. 
These point sources were selected from the second epoch image.
All pixels of the second epoch image were then set to zero except
the pixels of the selected point source of known magnitude 
and the resulting image was
then 
shifted and added to the first epoch image.
This analysis showed that we would have 
been able to detect (at 3 $\sigma$) a fading afterglow superimposed on 
this galaxy if its $I$-band magnitude had been 23.5 at the time of the 
first FORS2 observation.

\subsection{Upper limits on a rising supernova component}

One of the  main observational characteristics of a short burst should be the
absence of a SN component in the late-time afterglow \citep{Hjorth2005a}, as
the merger is not expected to result in the kind of radioactivity-powered
optical display typical for thermonuclear (Type Ia) and core-collapse (Types
II and Ib/c) supernovae. However, mergers may have sub-relativistic explosions
with low amount of ejected mass \citep{Li1998,Kulkarni2005}, but they should have 
a small luminosity. In agreement with these expectations, strong upper limits could be
set so far on any potential  SN component accompanying short bursts
\citep[cf.][]{Hjorth2005a, Fox2005}.

The constraints we can place on a rising SN component for GRB~050813 are less
severe, given the potentially relatively high redshift of this burst. For the
cosmological parameters employed here, SN 1998bw \citep{Galama1998} redshifted
to $z$=0.72 would have magnitudes of $I$=24.7 and $I$=23.9 during our first
and second VLT/FORS observing run, respectively, after taking into account a
Galactic reddening of $E(B-V)$=0.056 mag \citep{SFD1998} in the direction of
GRB 050813.  At that brightness level we would have detected the SN if not
superimposed on a much brighter host or strongly extinguished by dust. More
precisely, we conclude that at the time of our second FORS2 observation any
supernova following GRB 050813 was at least about 1.5 mag less luminous than
SN 1998bw. While constraints placed on any SN component underlying the
afterglow of e.g. GRB 050509B \citep{Hjorth2005a} and GRB 050709
\citep{Fox2005,Covino2006} are much stronger, this makes  a potential SN
component following GRB 050813 already fainter than any of the 11 GRB-SNe of
long bursts known to date \citep[ their Figure 6]{Ferrero2006}.

On the other hand, we would have been able to detect (at 3 $\sigma$) a rising
SN component superimposed on the bright galaxy E  (Fig.~\ref{field}) only if
its $I$-band magnitude had been 23.5 at the time of the second FORS2
observation. In other words, a SN 1998bw-like component would be missed in
this case. The same holds for a typical type Ia supernova
\citep{Krisciunas2003},  which would have had $I$=26.9 and $I$=25.4 at the
time of our first and second FORS2 observing run, respectively.

\section{Discussion}

Short bursts, by phenomenological classification introduced by
\cite{Kouveliotou1993}, are bursts whose $T_{90}$ duration measured with
\emph{BATSE} was less than 2 sec. Even though it has already been known in the
1990s that $T_{90}$ is a function of energy (and of detector properties),
this definition, because of its simplicity, has been widely used  even in the
\emph{HETE-2}  and in the \emph{Swift} era. In principle,  having now much
more observational data at hand for individual bursts than in the \emph{BATSE}
era, this phenomenological definition/classification scheme calls for a more
accurate, namely  physical classification scheme. 

It is clear that the classification of individual bursts with respect to the
nature of their progenitor is difficult. Recent investigations  tackle this
problem and have led to the suggestion of much more then just one criterium in
order to classify a GRB \citep{Donaghy2006,NB2006}. As long as no consensus
has been reached in the literature what the ultimate criteria are for a burst
to be classified as being due to a merger event, in several cases only
arguments can be provided that favor one scenario for the other (merger
vs. collapse). The detection or non-detection of a SN signal plays a key role
in this approach  but has come into question recently
\citep[see][]{Gehrels2006,Fynbo2006,DellaValle2006,Gal-Yam2006,Zhang2006}.
This leaves the nature of the host galaxy as the strongest  argument to detect
a GRB due to a merger event, namely  if the host is an elliptical galaxy. But
the potentially  broad range in merger times and hence distances of the merger
events from their host galaxies \citep[cf.][]{Belczynski2006} might also call
into question the application of this criterium. GRB 050813 belongs to those
bursts that demonstrate all these problems in detail.

One of the main goals of our observing runs was the localization of the
afterglow and hence the identification of the GRB host galaxy.  Basically, the
host cannot be identified with certainty  and we have to consider other
arguments that favor or disfavor any galaxy visible on the deep FORS2 $I$-band
images of the XRT error circle as the potential host. GRB 050813 then joins
the increasing list of short bursts with no detected optical afterglow,
starting with GRB 050509B \citep{Bloom2006a,CT2005, Gehrels2005,Hjorth2005a}.
Using the upper limits on the afterglow of GRB 050813 (Table 1) we can follow
\cite{Kann2006} and place the properties of this afterglow in the context of
other known GRB afterglows (Fig.~\ref{IcLKs}).  The long burst afterglows
shown in Fig.~\ref{IcLKs} by solid lines are those from the ``Golden Sample'' of
\cite{Kann2006}, i.e., those that have sufficient $I$-band data. In addition,
we analyzed the available afterglow  data on the short bursts GRB 050709
\citep{Hjorth2005b, Fox2005, Covino2006}, GRB 050724 \citep{Berger2005a,
Malesani050724}, GRB
051221A \citep{Soderberg2006} and  GRB 060121 \citep{Levan2006,dUP2006} in an
analogous way and also included them in Fig.~\ref{IcLKs} (see the Appendix B
for details). As can be seen, short burst optical afterglows  are
intrinsically very faint, with the afterglows of GRB 050724 and GRB 051221A
being about 3 magnitudes fainter than any long burst afterglow in the sample,
and GRB 050709 being 4 magnitudes fainter at one day after the burst and
assuming  $z=0.72$ \citep[in agreement with the predictions for short burst
afterglows;][]{Panaitescu2001}. They are also significantly fainter than
intrinsically faint afterglows  of some long GRBs, such as GRB 021211. Only
the afterglow of GRB 060121 is comparable with the typical afterglows of long
GRBs.   The upper limits on the optical afterglow of GRB 050813 show that its
luminosity was also far below typical luminosities of (extinction-corrected)
afterglows of long bursts. On the other hand, it matches the luminosity region
occupied so far by afterglows of the short bursts (with GRB 060121 being the
only exception).

\begin{figure}
\includegraphics[width=16.5cm,angle=0,clip=true]{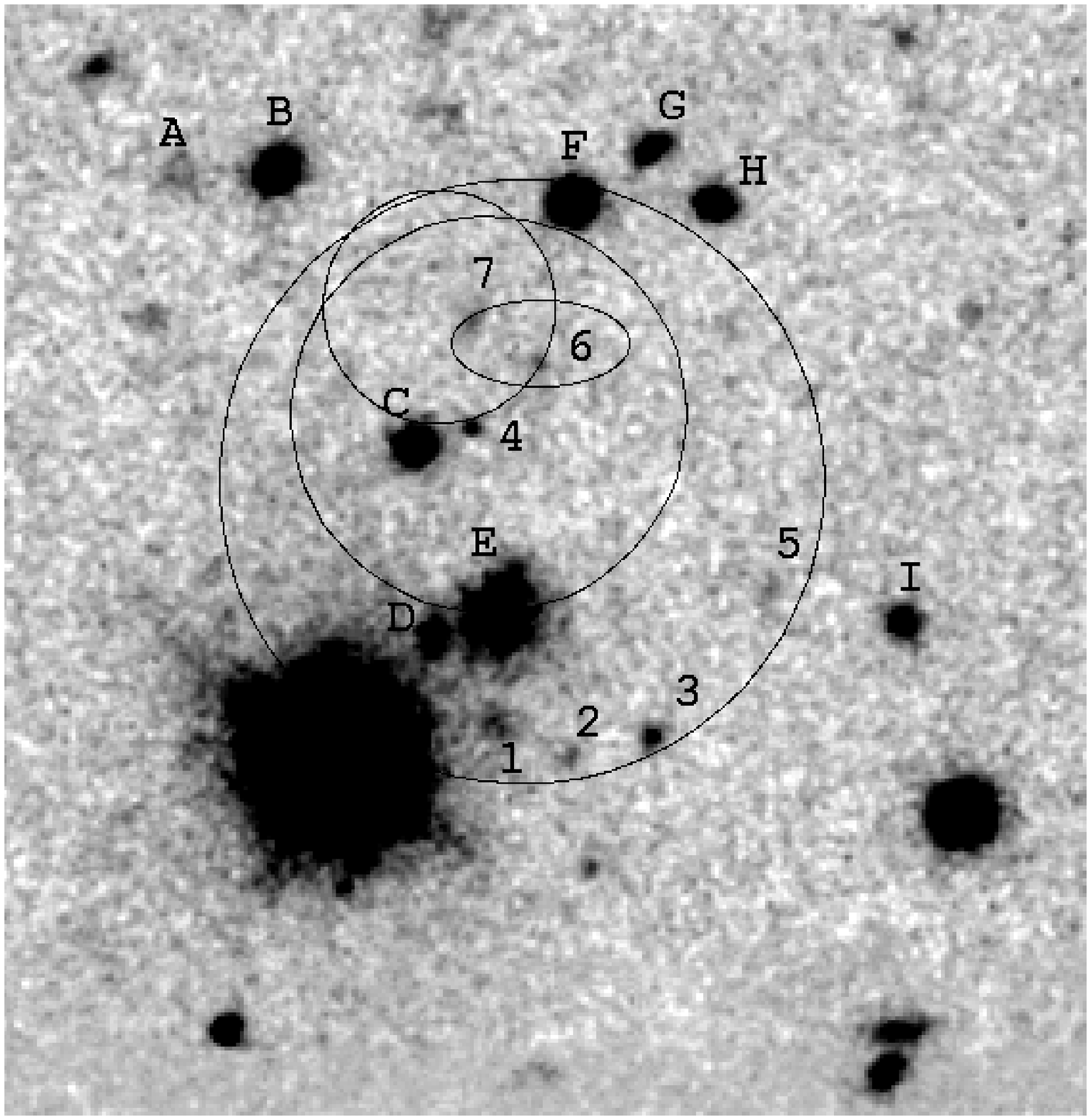}
\caption{VLT $I$-band image of the GRB field obtained  11 days after the
burst, showing the  original 10 arcsec (radius) XRT error circle of GRB 050813
\citep{Morris2005} (large circle), the refined error circle by
\cite{Moretti2006} (small circle, center around source \#4), the revised error
ellipse \citep{Prochaska2006}, the refined error circle by \cite{Butler2006}
(small circle, center around source \#7) and the objects listed in
Tables~\ref{tab1} and \ref{tab2}.}
\label{field}
\end{figure}

Figure~\ref{field} shows that there are only two sources in the XRT error
ellipse \citep{Prochaska2006}, while there are at least three additional
sources in the refined error circle \citep{Moretti2006}. The former might
favor a burst related to the very faint sources \#6 and \#7 (source \#6
appears  point-like in our images) but it does not even exclude an event in
the outer halo of source C, an elliptical galaxy at a redshift of 0.719
\citep{Prochaska2006}. The minimum distance between the border of the error
ellipse and the center of this galaxy is 3.2 arcsec, corresponding to a
projected distance of 23 kpc. This is less than the projected distance of the
error circle of GRB 050509B from the center of its suspected host, an
elliptical galaxy at a redshift of $z$=0.225 \citep{Gehrels2005}.  In
addition, the minimum angular distance between source E and the border of the
error ellipse is 7.1 arcsec, corresponding to a projected distance of 51
kpc. Even this is within the range predicted by recent models of merging
compact objects \citep[see][]{BBK02,PB02}.  The error circle determined by
\cite{Moretti2006} is much larger, and thus allows not only source C but also
galaxy E at $z=0.73\pm0.01$ \citep{Prochaska2006} to be the potential host of
GRB 050813. This galaxy was classified by \cite{Prochaska2006} as an elliptical
galaxy, while our images show morphology that point either to a spiral or to
an irregular galaxy. The nature of the fifth, point-like source in the refined
error circle, \#4, remains undetermined.

While this paper was submitted, a new revised XRT error circle was reported by
\cite{Butler2006}. This revised error circle is 3.8 arcsec in radius and
centered close to a faint  edge-on galaxy. This galaxy (source \#7, see
Fig.~\ref{field}) was only marginally detected during the first FORS
observations. A comparison with the second FORS observations six days later
does not provide convincing evidence for a photometric variability due to an
underlying point source.

To summarize, our optical data do not reveal either an afterglow nor a SN
component. If GRB 050813 was occurring in a cluster of galaxies at a redshift
of $z$=0.72, as it might be indicated by the surrounding galaxy population,
then its projected distance from its potential host galaxy could have been of
the order of less than 4 to some dozen kpc, depending on the chosen potential
host galaxy. The non-detection of the afterglow is well in accord with the
faintness of optical afterglows following short bursts (Fig.~\ref{IcLKs}). On
the other hand, if the burster would had been at $z$=1.8 \citep{Berger2006},
no SN~1998bw-like component would have been detectable  in our images and any
afterglow component would have been correspondingly fainter than in the former
case (Fig.~\ref{IcLKs2}). But even in this case  the upper limits we can set
on any optical afterglow are consistent with the hypothesis that GRB 050813
was a typical member of the short bursts.

\section{Acknowledgements}

We thank the staff at ESO/Paranal, in particular C. Dumas, P. D. Lynam,
P. Gandhi, N. Hu\'elamo, and E. Jehin, for performing the observations and
additional efforts related to that. We thank CAHA and OSN staff for excellent
support during some of the observations  presented here.  P.F., D.A.K., and
S.K. acknowledge financial support by DFG grant Kl 766/13-2 and by the German
Academic Exchange Service (DAAD) under grant No. D/05/54048. The research
activity of J.Gorosabel is supported by the Spanish Ministry of Science
through projects AYA2004-01515 and ESP2005-07714-C03-03. We thank the  second
referee for a rapid reply and a constructive report.

\appendix

\section{The light curves of the short burst afterglows}

\begin{figure}
\includegraphics[width=16.5cm,angle=0,clip=true]{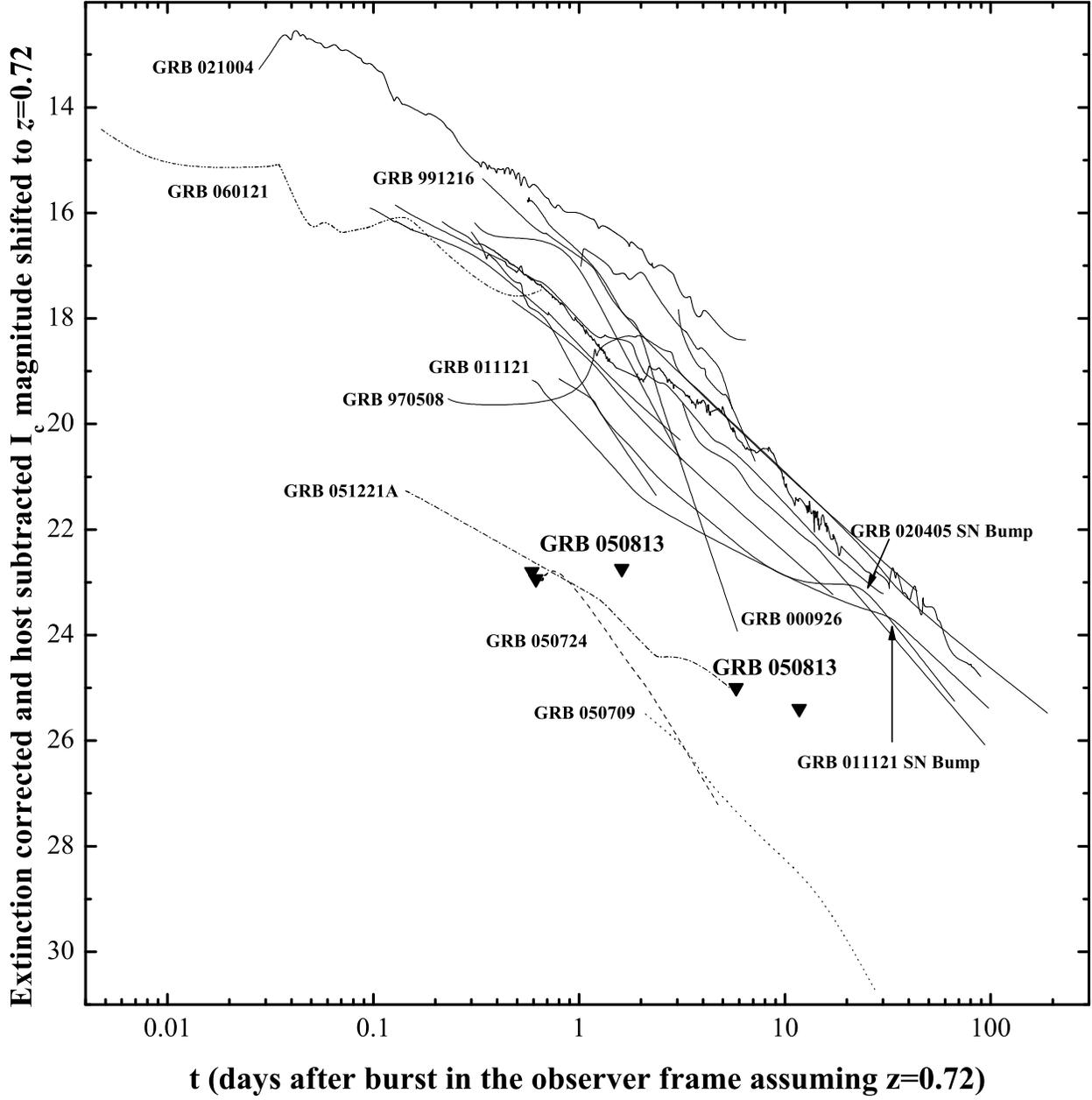}
\caption{The $I$-band light curves of all afterglows from the ``Golden Sample''
of  \cite{Kann2006} after correction for Galactic and for host extinction and
after shifting them to a common redshift of $z$=0.72, the potential redshift
of GRB 050813. Two long GRB supernova rebrightenings are indicated. Also shown
are the $I$-band afterglows of the short bursts GRB 050709, 050724, 051221A
and 060121 shifted in a similar way, and our upper limits on any afterglow or
supernova from GRB 050813 (upside-down triangles).  For GRB 060121 a redshift
of $z=4.6$ \citep{dUP2006} is assumed here.}
\label{IcLKs}
\end{figure}

\begin{figure}
\includegraphics[width=16.5cm,angle=0,clip=true]{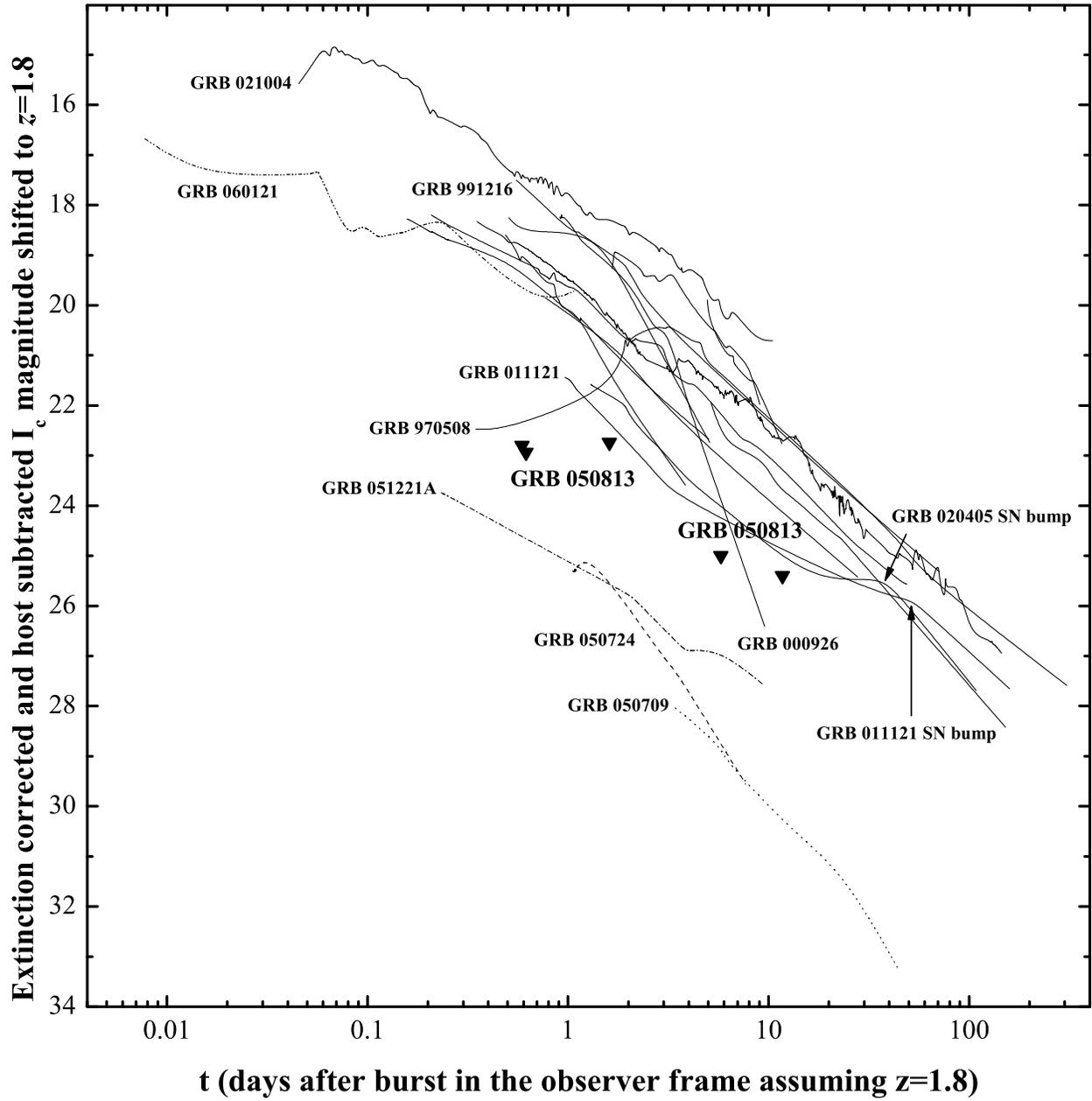}
\caption{The same as Fig.\ref{IcLKs}, but for a redshift of 1.8}
\label{IcLKs2}
\end{figure}

In Fig. \ref{IcLKs} we included those four GRBs that have both an optical
afterglow and a redshift derived either from host galaxy spectroscopy or
photometry \citep[GRB 060121;][]{dUP2006} up to October 2006.

We take data from the following works: GRB 050709: \cite{Hjorth2005b,
Fox2005, Covino2006}. GRB 050724: \cite{Berger2005a, Malesani050724}.
GRB 051221A: \cite{Soderberg2006}. GRB 060121: \cite{Levan2006, dUP2006}.

For GRB 050709, we derive a decay slope of $\alpha=1.68\pm0.15$ from the
$R_C$-band light curve. \cite{Fox2005} noted that the late \emph{Hubble Space
Telescope (HST)} data indicate a steepening of the light curve decay, possibly
due to a jet break. Using the $R_C$-band decay index, we find a rebrightening
(significant at the 5 $\sigma$ level) in the HST data, but only marginal
evidence that the afterglow is fainter than expected from the early decay in
the last HST detection. This result is in accordance with \cite{Watson2006}.
The light curve shown in Fig.~\ref{IcLKs} is composed of the $R_C$ data
shifted to the HST F814W zero point, plus the HST data. From the
$V,R_C,F8,K^\prime$ spectral energy distribution (SED), we derive a steep
uncorrected spectral slope $\beta_0=1.71\pm0.17$. This is indicative of
additional source frame extinction. As the host is a blue dwarf galaxy
\citep{Fox2005}, we assumed SMC-type dust \citep{Pei1992}. A free fit implies
$\beta=0.26\pm1.16$ and a host extinction  of $A_V$(host)$=1.46\pm1.07$ mag, a
very high value indeed. As the single $K^\prime$-data point has a very large
error (0.7 mag), this value may not be trustworthy. For a progenitor that has
traveled far from its birthplace, an unstratified surrounding medium is
expected (density $\rho \propto r^0$). We fixed $\beta$ to the value derived
from the pre-break decay slope $\alpha_1$, and find $\beta=1.12$ and
$A_V$(host)$=0.67\pm0.19$ mag. We used these parameters to correct and shift
the light curve.

For GRB 050724, the Galactic extinction is high and not well determined.
We follow \cite{Malesani050724}, who argue, based on the X-ray to optical SED,
for $E_{B-V}=0.49$. After correcting for this extinction, we find $\beta=0.76
\pm0.07$ and no evidence for source frame extinction, in accordance with
\cite{Malesani050724}. The light curve is mostly $I_C$ data anyway, we
add $V$, $R_C$ and $K$ data shifted to the $I_C$ zero point.

In the case of GRB 051221A, we find that the light curve decays as a single
power-law with a slope $\alpha=0.94\pm0.03$, in accordance with
\cite{Soderberg2006}.  We derive a flat spectral slope ($\beta=-0.16\pm0.84$)
from the $r^\prime i^\prime z^\prime$ spectral energy distribution, but
caution that the errors of the $i^\prime$ and $z^\prime$ data are very
large. Assuming an unstratified surrounding medium and a cooling frequency
blueward of the optical bands, we derive $\beta=0.62$ \citep[coupled with a
typical  power-law index of the electron distribution function of $p=2.25$;
cf.][]{Kann2006}. We used this spectral slope and assume no additional
extinction to shift the light curve.

Combining the data from \cite{Levan2006} and \cite{dUP2006} of GRB 060121, we
find that the zero points of the two data sets differ. We shifted the data
from \cite{dUP2006} to the fainter zero point of \cite{Levan2006}. The light
curve has a complex shape and seems to include several rebrightenings
(Fig. \ref{IcLKs}). It is composed of $I_C$ data and $R_C$ data shifted to the
$I_C$ zero point. We used the redshift and host galaxy extinction derived by
\cite{dUP2006}, assuming the more probable redshift of $z=4.6$, and a
spectral slope in the optical of $\beta=0.6$, as derived by the authors cited
above.

In all cases, except for GRB 060121, the afterglow data do not contain any
host contribution. For GRB 060121, we used a host galaxy magnitude derived
from the HST measurements \citep{Levan2006}. To correct for
Galactic extinction, we used the value derived from the maps of \cite{SFD1998}
for GRB 050709, 051221A and 060121, and $E_{B-V}=0.49$ mag for GRB 050724 
\citep[as suggested by][]{Malesani050724}.



\clearpage

\begin{deluxetable}{lccccc}
\tablecolumns{6}
\tabletypesize{\scriptsize}
\tablecaption{Observing log of the GRB 050813 field}
\tablehead{
\colhead{Date [days]} &
\colhead{$t-t_0$\tablenotemark{a} [days]} &
\colhead{Mag\tablenotemark{b}} &
\colhead{Exposure [s]} &
\colhead{Filter} &
\colhead{Telescope}}
\startdata
13.8333 & 0.5519 & 22.8 &10$\times$600 &I  &  1.5m OSN \\
13.8708 & 0.5894 & 23.3 &23$\times$180 &R  &  2.2m, CAFOS\\
14.8475 & 1.5661 & 23.1 &24$\times$300 &R  &  2.2m, CAFOS\\
19.0606 & 5.7792 & 25.1 &10$\times$200 &I  &  8.2m, FORS2\\
24.9901 &11.7087 & 25.5 &10$\times$200 &I  &  8.2m, FORS2\\ \hline
\enddata
\tablenotetext{a}{$t_0$ = 2005 August 13.2814, the time of the burst. 
All dates refer to August 2005 and give the time of
the start of the first exposure.}
\tablenotetext{b}{The limiting magnitude of the combined image.}
\label{log}
\end{deluxetable}


\newpage\clearpage

\begin{deluxetable}{cccc}
\tablecolumns{4}
\tabletypesize{\scriptsize}
\tablecaption{The objects used for the calibration of the photometry
(A,B,F,G,H,I) and the brightest galaxies in the XRT error circle (C,D,E).}
\tablehead{
\colhead{\#\tablenotemark{a}} &
\colhead{RA\tablenotemark{b}} &
\colhead{DEC\tablenotemark{b}} &
\colhead{$I$}}
\startdata
A      & 16:07:57.72 & +11:15:02.24 & $24.68\pm0.35$ \\
B      & 16:07:57.50 & +11:15:02.13 & $21.83\pm0.09$ \\
C      & 16:07:57.19 & +11:14:53.15 & $22.43\pm0.12$ \\
D      & 16:07:57.16 & +11:14:46.86 & $23.38\pm0.22$ \\
E      & 16:07:57.01 & +11:14:47.61 & $22.74\pm0.28$ \\
F      & 16:07:56.85 & +11:15:01.80 & $20.88\pm0.03$ \\
G      & 16:07:56.66 & +11:15:02.87 & $23.61\pm0.19$ \\
H      & 16:07:56.53 & +11:15:01.11 & $22.85\pm0.14$ \\
I      & 16:07:56.10 & +11:14:47.34 & $23.50\pm0.17$ \\
\enddata
\tablenotetext{a}{The numbering follows Fig.~\ref{field}.}
\tablenotetext{b}{Epoch J2000}
\label{tab1}
\end{deluxetable}

\newpage\clearpage

\begin{deluxetable}{ccccc}
\tablecolumns{4}
\tabletypesize{\scriptsize}
\tablecaption{The photometry of the fainter sources in the XRT error circle.}
\tablehead{
\colhead{\#\tablenotemark{a}} &
\colhead{RA\tablenotemark{b}} &
\colhead{DEC\tablenotemark{b}} &
\colhead{$I$ run 1\tablenotemark{c}} &
\colhead{$I$ run 2\tablenotemark{c}}}
\startdata
1      & 16:07:57.00 & +11:14:43.83 & $24.7<I<24.9$     & $24.4<I<25.4$   \\
2      & 16:07:56.85 & +11:14:42.91 & $>25.1$            & $24.4<I<25.5$   \\
3      & 16:07:56.66 & +11:14:43.58 & $24.69\pm0.24$   & $24.44\pm0.10$  \\
4      & 16:07:57.07 & +11:14:53.65 & $24.63\pm0.30$   & $24.67\pm0.13$  \\
5      & 16:07:56.40 & +11:14:48.35 & $>25.1$           & $25.47\pm0.25$  \\
6      & 16:07:56.91 & +11:14:55.91 & $>25.1$           & $25.64\pm0.28$  \\
7      & 16:07:57.07 & +11.14.57.43 & $24.7<I<25.1$    & $25.41\pm0.25$  \\
\enddata
\tablenotetext{a}{The numbering follows Fig.~\ref{field}.}
\tablenotetext{b}{Epoch J2000}
\tablenotetext{c}{Run 1 and run 2 refer to the first
and second VLT/FORS observations, respectively.}
\label{tab2}
\end{deluxetable}

\end{document}